\documentclass[reprint,superscriptaddress,amsmath,amssymb,aps]{revtex4-2}
\usepackage[utf8]{inputenc}
\usepackage[T1]{fontenc}
\usepackage{graphicx}
\usepackage{dcolumn}
\usepackage{bm}
\usepackage{float}
\usepackage{multirow}
\usepackage{amsfonts}
\usepackage{subcaption}
\usepackage{booktabs}
\usepackage{url} 
\usepackage[hidelinks]{hyperref} 
\usepackage{orcidlink}
\usepackage{multirow}
\usepackage{booktabs}

\begin{document}
	
	\title{Cortical information transfer reveals conserved hemispherical network dynamics across human handedness}
	
	\author{Yago Emanoel Ramos \orcidlink{0009-0003-7125-8199}}
	\email{yago.emanoel@ufba.br}
	\affiliation{Biosystems Laboratory, Instituto de Física, Universidade Federal da Bahia, Salvador, Brazil}
	\affiliation{Department of Theoretical Physics, Faculty of Sciences, University of Zaragoza, Zaragoza, Spain}
	\affiliation{Institute for Biocomputation and Physics of Complex Systems (BIFI), University of Zaragoza, Zaragoza, Spain}

	\author{José Garcia Vivas Miranda \orcidlink{0000-0002-7752-8319}}
	\email{vivas@ufba.br}
	\affiliation{Biosystems Laboratory, Instituto de Física, Universidade Federal da Bahia, Salvador, Brazil}

	\begin{abstract}
		
		Whether human motor and brain lateralization arises from fundamentally distinct neural architectures or emerges from conserved network dynamics remains a central question at the intersection of network science and neurobiology. Conventional measures of cortical activation often fail to resolve how directed information exchange adapts to manual preference during complex motor tasks. This ambiguity leaves it unclear whether left-handed individuals possess atypical neural organization or follow shared dynamical principles. Here, we combine Permutation Transfer Entropy (PTE) with network-based indices to map directed cortical signal flow using electroencephalography (EEG) during high-precision motor execution (handwriting) in both right- and left-handed individuals. We demonstrate that cortical communication is strictly scale-dependent regardless of handedness: ipsilateral dominance occurs primarily in general, network-wide interactions, whereas inter-hemispheric interactions maintain a contralateral profile. Crucially, left-handed individuals exhibit inter-hemispheric information dynamics that are functionally equivalent to right-handers when executing tasks with the same anatomical hand, refuting assumptions of mirrored or atypical lateralization. By demonstrating that human cortical lateralization relies on dynamical principles conserved across manual preferences, these findings suggest that functional asymmetries reflect adaptive optimizations governed by epigenetic or social flexibility, rather than fixed structural divergences. 
	
	\end{abstract}

	\maketitle

	\date{}
	\section{Introduction}
	Laterality is a fundamental property of biological systems, observable across taxa ranging from single-celled organisms to the blue whale. It is defined by structural or functional asymmetries between the bilateral halves of the body~\cite{Wiper02112017}. While morphological symmetry is deeply conserved in biology, distinct evolutionary trade-offs, both advantageous and disadvantageous, emerge in systems that develop specialized lateralized functions~\cite{blum_animal_2018,Ocklenburg2022}.
	
	Neuro and motor behavior are deeply connected~\cite{williams_structural_2023, ocklenburg_genetics_2025, RAMOS2026118081, doi:10.1177/15500594251394773} and in the context of human motor neurodevelopment, laterality is exceptionally pronounced~\cite{10.7554/eLife.77875}. \textit{Homo sapiens} exhibits the highest known degree of population-level handedness, characterized by a strong right-hand dominance~\cite{Verendeev2016}. While many quadrupedal species lack clear evolutionary pressures for manual specialization, primates evolved highly lateralized upper-limb functions~\cite{https://doi.org/10.1111/eth.12827}. This includes the execution of fine motor tasks coupled with impedance control, which is typically provided by the non-dominant hand~\cite{sainburg_evidence_2002,renfrew_neural_2008,10.3389/fpsyg.2014.01092,YADAV2014385}. The advanced specialization and stabilization of fine motor movements represent a cornerstone of human cognitive and physical evolution, enabling critical adaptations such as the manipulation of fire and the fabrication of complex tools for logistics and hunting~\cite{10.3389/fpsyg.2017.01021}.
	
	Cerebral hemispheres typically operate contralaterally, with the left motor cortex primarily governing fine motor execution in the right hand~\cite{corballis_evolution_2008, Sha2021}. Despite the evolutionary triumph of Homo sapiens and our extraordinary dexterity, approximately $10\%$ of the population (left-handers) exhibits inverted manual dominance with comparable motor performance~\cite{RAMOS2025103425,RAMOS2025117412}. Notably, left-handed individuals are often disproportionately represented at the elite levels of interactive sports, sometimes comprising up to half of the athletes in specific competitive modalities, suggesting distinct performance selection mechanisms associated with non-standard handedness~\cite{hagemann_advantage_2009,loffing_left-handedness_2017,simon_prevalence_2025}. Furthermore, laterality is not strictly genetically determined; the existence of monozygotic twins with discordant handedness~\cite{BADZAKOVATRAJKOV20103086, steinmetz_brain_1995} highlights the complex interplay between genetic, epigenetic, and environmental factors. This raises the question of how differences in cortical organization between right- and left-handers may emerge through environmental and social influences rather than being predetermined. Consistent with this view, Klöppel et al.~\cite{kloppel2007can} showed that left-handers forced to use their right hand during childhood undergo cortical reorganization toward a pattern more closely resembling that of right-handers, suggesting that neural differences associated with handedness are shaped by experience.
	
	This behavioral divergence poses a fundamental evolutionary puzzle. If handedness is so deeply embedded in human motor organization, why some individuals consistently rely on the opposite hand without substantial functional impairment? The answer may lie in how the brain organizes motor function rather than in which hand is preferred. In right-handers, hand dominance and hemispheric specialization are closely aligned~\cite{TZOURIOMAZOYER2016319}; in left-handers, they can become uncoupled~\cite{KARLSSON2024108837}. Thus, the critical question is whether the brain follows the anatomical hand or its functional dominance. In simple movement and motor imagery tasks, there is evidence of mirrored or attenuated neural responses in left-handers~\cite{10.3389/neuro.09.039.2009,10.1371/journal.pone.0036036}. However, these studies are limited by tasks that do not impose substantial demands for neuromotor reorganization, such as handwriting, which subjects naturally perform with their dominant hand, and by approaches that reduce cerebral lateralization to interactions between electrodes C3 and C4~\cite{FERNANDES2027116449}, thereby overlooking the complex network of interactions inherent to brain motor function~\cite{feng_multimodal_2026}.

	Previous studies in handwriting utilizing metrics such as spectral power~\cite{RAMOS2025103425} and permutation entropy~\cite{RAMOS2025117412} have demonstrated that within-subject neuromotor differences between hands are significantly larger in right-handers than in left-handers. However, between-subject comparisons (right-handers versus left-handers) fail to yield significant differences, regardless of whether the hands are grouped by functional role (dominant versus non-dominant) or anatomical side (right versus left). This discrepancy leaves a critical ambiguity: it remains unclear whether unpaired analyses are simply confounded by high inter-individual variance, or if identifying the left-handed neural equivalent of a right-hander's dominant grasp requires more sophisticated metrics sensitive to the directionality of information.

	To investigate these questions, we move beyond regional activation and examine how cortical regions exchange information during complex motor behavior. Conventional measures of spectral power and (de)synchronization characterize neural activity but provide limited insight into the direction of cortical interaction. Here, we use Permutation Transfer Entropy (PTE) to quantify directed information transfer across cortical regions, asking whether the architecture of motor communication changes with handedness. By comparing neural dynamics across the same anatomical hand in left- and right-handers, this study seeks to uncover a fundamental principle of human motor lateralization: whether the brain's organization is ultimately shaped more strongly by the hand itself, by its functional dominance, or by the demands of behavior. We therefore test whether handedness alters the fundamental structure of cortical communication, or primarily its level of lateralization.

	\section{Results}
	
	\subsection{Experimental Design}
	
\begin{figure*}
		\centering
		\includegraphics[width=0.9\textwidth]{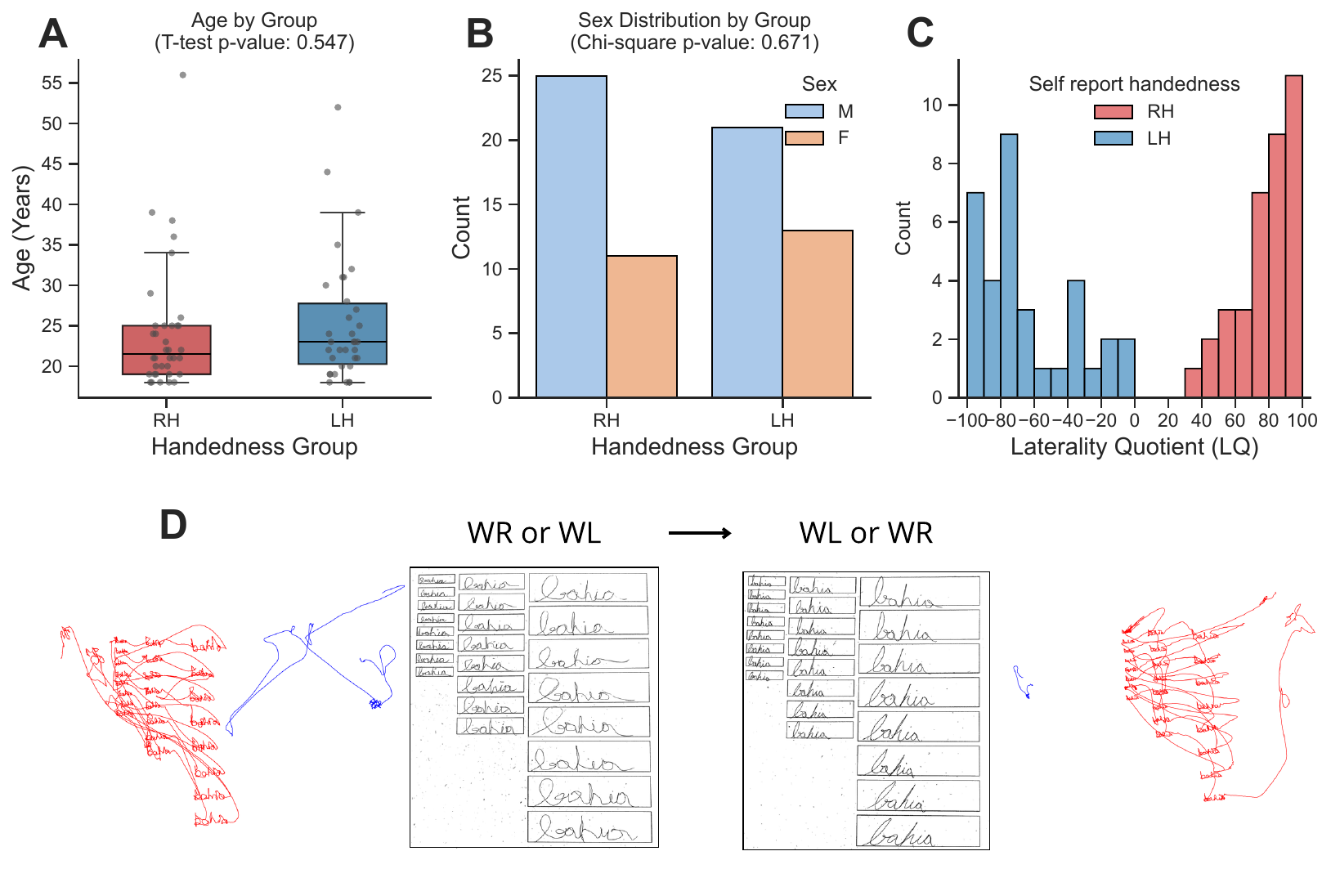}
		\caption{\textbf{Sample demographics and experimental setup.} (A) Age distribution across the right-handed (RH) and left-handed (LH) groups, showing no significant difference (independent t-test, p = 0.547). (B) Biological sex distribution per group, indicating matched proportions (Chi-square test, p = 0.671). (C) Distribution of Laterality Quotients (LQ) derived from the Edinburgh Handedness Inventory (EHI) for self-reported RH and LH participants. (D) Experimental handwriting output example. Participants performed writing tasks with both their right (WR) and left (WL) hands in a randomized order, filling eight boxes of three specific sizes in a A4 sheet. The adjacent spatial trajectories illustrate the trajectory differences between executing the task with the non-dominant versus dominant hand. Red traces represent the active (executing) hand, while blue traces depict the auxiliary hand (providing impedance control).}
		\label{Fig1}
	\end{figure*}

	In this study, we employed a 64-channel, 1000hz electroencephalography (EEG) collection during handwriting task in which subjects were instructed to repeatedly write the word `bahia' (the name of both the state and the university where the experiment was conducted). This specific word was chosen to minimize linguistic confounding factors, as it is a highly familiar, everyday word for all participants. All subjects performed the task using each hand separately, in a randomized order (Fig.~\ref{Fig1}D). No time limit was set for the task, allowing participants to perform it at their own pace. However, for the analysis, only the first minute of the EEG recording was used in order to preserve the temporal characteristics of the electrodes interaction.
	
	The experiment was conducted with volunteers from the university community (including students, technical staff, and professors), resulting in a total sample of 70 individuals. Based on self-declaration, the cohort comprised 36 right-handers and 34 left-handers. Fig.~\ref{Fig1}A presents the age distribution of the two groups; an independent t-test revealed no significant difference in age (p = 0.568). Similarly, Fig.~\ref{Fig1}B demonstrates that the sex distribution was not significantly different between the groups (Chi-square p = 0.859). Five subjects (2 right-handed and 3 left-handed) were excluded from this study due to the presence of excessive muscle artifacts.
	
	Fig.~\ref{Fig1}C illustrates the distribution of the Laterality Quotient (LQ) from the Edinburgh Handedness Inventory test, which quantifies individuals' propensity to use the left hand (negative values) or the right hand (positive values). It is important to note that while the LQ test scores might classify some subjects as ambidextrous, previous studies indicate that these individuals still exhibit equivalent lateralization during tasks requiring sufficiently high fine motor control~\cite{RAMOS2025117412}. Since no participant reported a lack of hand preference for fine motor tasks (such as writing, drawing, or sewing), we interpreted any ambidextrous tendencies in the LQ as flexibility restricted to gross motor tasks. Consequently, we classified the participants' handedness based on their self-reported preferences.

     \subsection{Global Connectivity and Hand Dominance}
     
     To evaluate the directional flow of information between brain regions, we used normalized PTE. Normalized by total entropy, $PTE_{x \rightarrow y}$ quantifies the empirical information transfer by representing the fraction of signal $y$'s information that becomes predictable when signal $x$ is known. By computing this index for all electrode pairs, we modeled the cortex as a network using graph-theoretic properties. Specifically, we analyzed the out-degree (total outgoing information originating from an electrode) and the in-degree (total incoming information received from the rest of the network). Throughout the analysis and visualization, groups (RH: right-handers, LH: left-handers) and tasks (WR: writing with the right hand, WL: writing with the left hand) are denoted as Group$\_$Task combinations (e.g., RH$\_$WR).

     \begin{figure*}  
     	\centering
     	\includegraphics[width=0.95\textwidth]{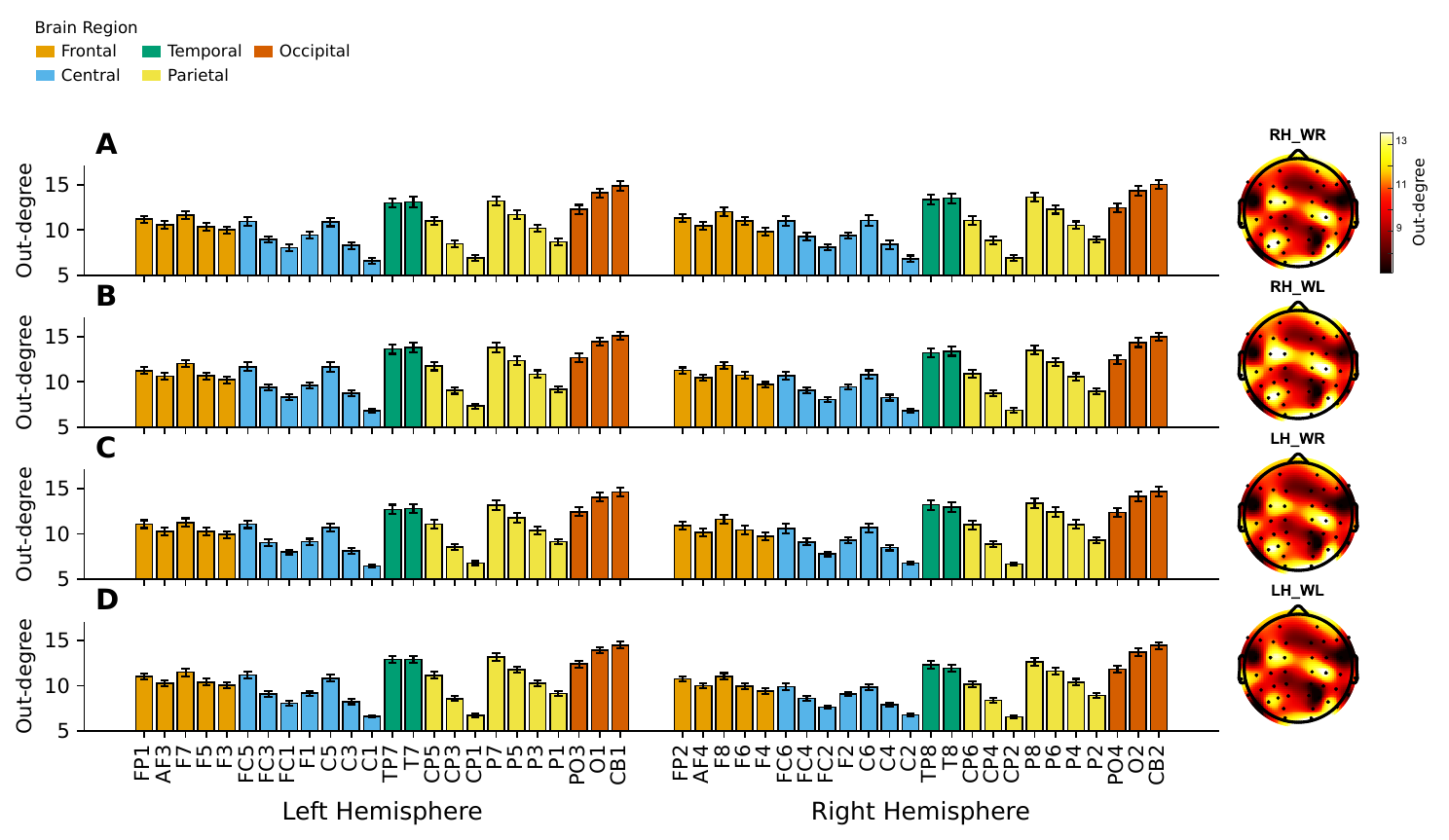}
         \caption{\textbf{Spatial distribution of out-degree.} Representing the mean extent to which each electrode influences all the others, shown as bar plots, with error bars representing the standard error across individuals, and brain topoplots for the four group$\_$task combinations: RH\_WR (A), RH\_WL (B), LH\_WR (C), and LH\_WL (D). For the topoplots, spatial interpolation between electrode sites was performed using biharmonic spline interpolation (default EEGLAB method) mapped onto a 2D circular grid.}
     	\label{Fig2}
     \end{figure*}

\begin{figure*}  
     	\centering
     	\includegraphics[width=1.0\textwidth]{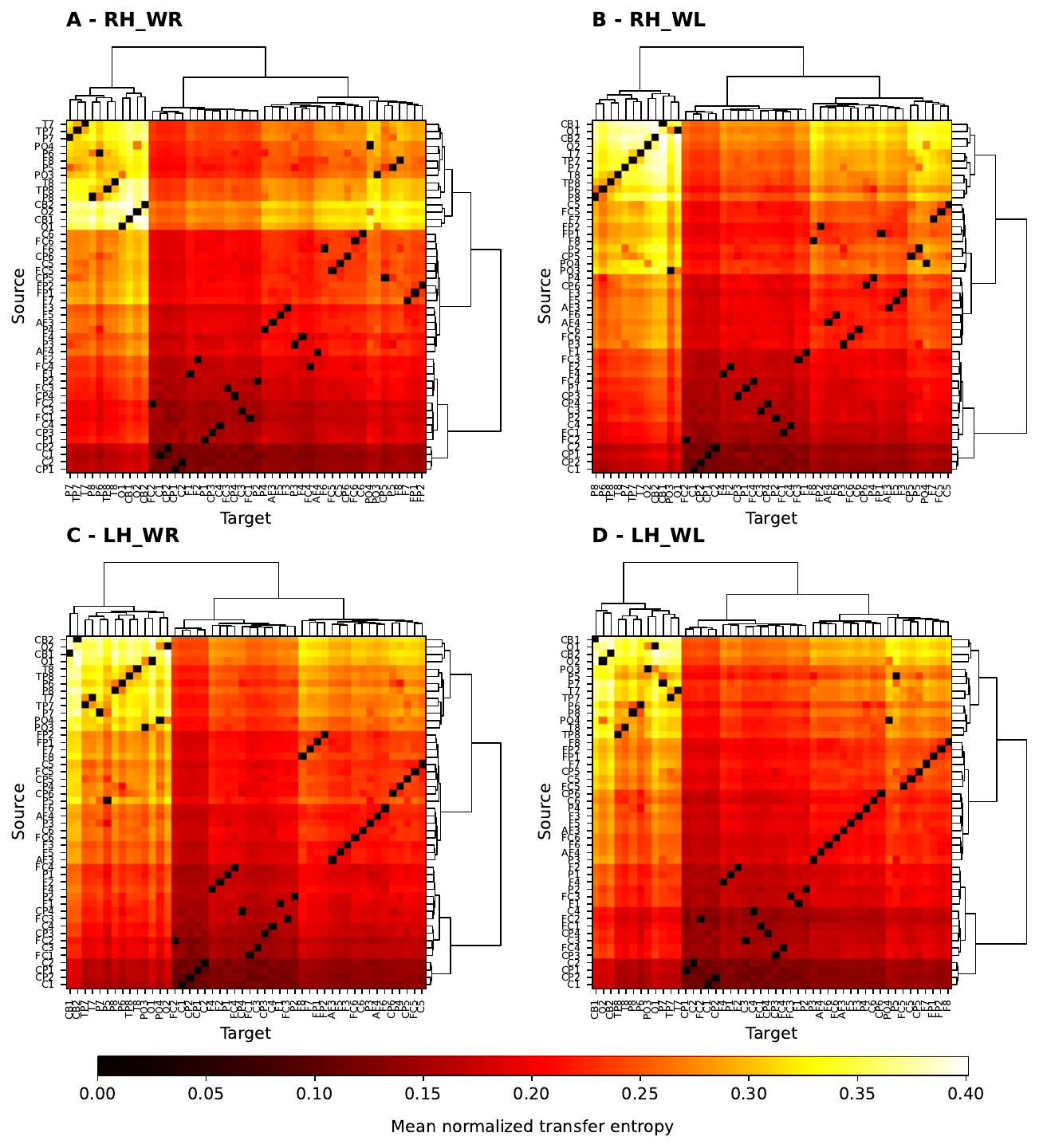}
     	\caption{\textbf{Average PTE hierarchical organization of adjacency matrices across groups and tasks.} The matrices show the fraction of the Target eletrode information that is explained by knowing the Source electrode. After averaging by all the subjects, the Source and Target electrodes were grouped and clustered using dendrograms based on Wards method: RH$\_$WR (A), RH$\_$WL (B), LH$\_$WR (C), and LH$\_$WL (D).}
     	\label{Fig3}
     \end{figure*}

      The out-degree analysis revealed that both tasks and handedness groups exhibit similar distributions with narrow inter-subject error margins. Localized increases in degree were observed exclusively within specific foci, namely the occipital, left central, and right parietal regions, across the different tasks (Fig.~\ref{Fig2}). To visualize this average incoming and outgoing flow on a network scale, Fig.~\ref{Fig3} displays the average source and target patterns across tasks. Notably, these overarching macro-scale patterns share a comparable hierarchical dendrogram structure. By focusing on the subtle variations and directionality patterns within the individual data, we can accurately characterize task- or group-specific differences and assess the consistency of these network patterns.

     \subsection{Pairwise individual comparisons and lateralization (a)symmetry}
     
     To understand how the hierarchical organization of PTE matrices (Fig.~\ref{Fig3}) varies between hands, we first evaluated the data at the individual level. We considered PTE matrices for the right and left hands of each subject. Next, we calculated the intra-subject pairwise differences and correlations between these two matrices for every individual. To quantify these individual inter-manual differences, we applied three metrics. First, we used the Mean Absolute Error (MAE) to calculate the average cell-by-cell difference between an individual's two matrices. Second, we applied the Mantel test, analogous to a Pearson correlation for matrices, to evaluate their structural association. Finally, we used the cophenetic correlation to assess the similarity of the hierarchical clustering patterns observed in Fig.~\ref{Fig3}, applying this to both the source and target hierarchies (Table~\ref{Table1}).
     
     This procedure yielded a distribution of individual inter-manual differences for each group. When comparing these distributions between right- and left-handed individuals, the mean values of the two groups were strikingly similar, featuring narrow error margins and tight 95$\%$ confidence intervals. Consequently, the inter-group comparisons yielded non-significant p-values (Mann-Whitney U test, see Table~\ref{Table1}), indicating that the magnitude of difference between a person's left and right hand is equivalent regardless of whether they are right- or left-handed.
     
     However, the core question of this study remains: is this lateralization equivalence driven by functional roles or anatomical constraints? This leads to two  hypotheses:
     \begin{itemize}
     	\item The two hands generate distinct neural patterns because they perform different functional roles (dominant vs non-dominant hands).
     	\item The two hands generate distinct neural patterns because they are distinct anatomical limbs (right vs left hands).
     \end{itemize}
     To resolve this, it is necessary to move beyond pairwise comparisons and explore connectivity patterns and network indices that enable robust inter-subject analyzes.
     
     \begin{table*}[htbp]
     	\centering
     	
     	\label{tab:network_comparisons}
     	\begin{tabular}{llccc}
     		\toprule
     		\textbf{Metric} & \textbf{Group} & \textbf{Mean $\pm$ SE} & \textbf{95\% CI} & \textbf{\textit{p}-value} \\
     		\midrule
     		\multirow{2}{*}{Mean Absolute Error (MAE)} & RH & 0.0252 $\pm$ 0.0032 & [0.0187, 0.0317] & \multirow{2}{*}{0.4452} \\
     		& LH & 0.0232 $\pm$ 0.0022 & [0.0187, 0.0277] & \\
     		\midrule
     		\multirow{2}{*}{Mantel correlation} & RH & 0.9639 $\pm$ 0.0048 & [0.9532, 0.9722] & \multirow{2}{*}{0.2578} \\
     		& LH & 0.9597 $\pm$ 0.0063 & [0.9455, 0.9702] & \\
     		\midrule
     		\multirow{2}{*}{Cophenetic correlation (Source electrode hierarchy)} & RH & 0.7361 $\pm$ 0.0269 & [0.6789, 0.7844] & \multirow{2}{*}{0.4608} \\
     		& LH & 0.7468 $\pm$ 0.0328 & [0.6758, 0.8042] & \\
     		\midrule
     		\multirow{2}{*}{Cophenetic correlation (Target electrode hierarchy)} & RH & 0.8892 $\pm$ 0.0160 & [0.8536, 0.9164] & \multirow{2}{*}{0.4921} \\
     		& LH & 0.8912 $\pm$ 0.0200 & [0.8455, 0.9239] & \\
     		\bottomrule
     	\end{tabular}
     	\caption{Comparison of intra-subject inter-manual PTE matrix metrics between right- and left-handed groups. Pairwise differences (MAE), matrix associations (Mantel test), and hierarchical topological correlations (cophenetic correlation) were calculated between the two hands for each participant. Group distributions were compared using Mann-Whitney U tests.}
     	\label{Table1}
     \end{table*}

     \subsection{Hemispheric Information Balance and External-Internal (EI) Index}
     
     To quantify functional asymmetries in interhemispheric communication, we defined the Flow Balance Index ($B$-index), which captures the net directional signaling of each hemisphere as the normalized difference between total outgoing and incoming information. A positive $B$-index indicates that a hemisphere acts predominantly as a net driver (broadcasting more directional signals than it receives), whereas a negative value reflects a net receiver or integrator.

\begin{figure}
     	\centering
     	\includegraphics[width=0.48\textwidth]{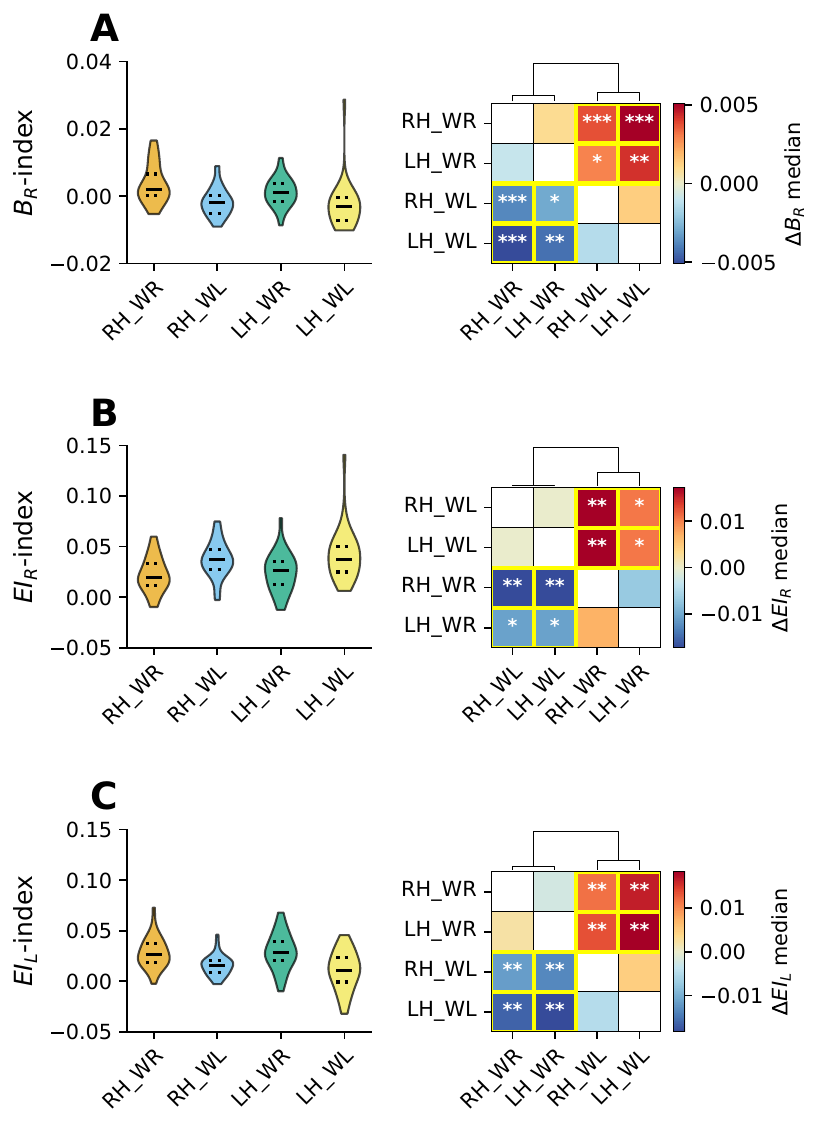}
     	\caption{\textbf{Distributions and pairwise dissimilarities of directional information flow indices.} Violin plots (left column) displaying the distribution of network indices across group-task combinations, alongside their corresponding dissimilarity matrices (right column) ordered by hierarchical clustering (dendrograms). Heatmap colors represent the difference in medians ($\Delta$ median) between conditions. Pairwise statistical comparisons were conducted using Mann-Whitney U tests, with Bonferroni correction applied for multiple comparisons. Significant differences are highlighted with asterisks (*p<0.05, **p<0.01, ***p<0.001). The analyzed metrics are: (A) Right hemisphere balance of information flow ($B_R$-index); (B) Right hemisphere external-internal flow ($EI_R$-index) and (C) Left hemisphere external-internal flow ($EI_L$-index).}
     	\label{fig4}
     \end{figure}

     Moving beyond individual pairwise hand comparisons, our analysis of interhemispheric network dynamics revealed that executing tasks with the left hand induces a net influx of neural signaling into the right hemisphere ($B_R > 0$). Because this index measures the complementary information balance between the two cerebral halves, the value for the left hemisphere is intrinsically inverted ($B_L = -B_R$). Crucially, these interhemispheric dynamics demonstrate that an individual’s trait handedness exerts a negligible influence on macro-scale cortical information balance. Instead, signal flow is overwhelmingly driven by the active anatomical limb. Non-parametric statistical comparisons (Mann-Whitney $U$ test) confirmed a robust divergence between tasks performed with different hands, but showed no significant difference between right- and left-handed individuals (Fig.~\ref{fig4}A).

     Similarly, the relative difference between external and internal connections (EI-index, measuring how heterophilic are the connections of each hemisphere) for the right hemisphere across all conditions exhibited a positive median during left-hand tasks and values near zero during right-hand tasks. Conversely, the left hemisphere displayed predominantly positive EI-index values for right-hand tasks and values near zero for left-hand tasks. This highlights a contralateral behavior in hemispheric integration: the hemisphere contralateral to the active executing hand exhibits greater external integration (Fig.~\ref{fig4}B,C).

     \subsection{Lateralization dominance (Ipsilateral vs. Contralateral)}

     Although the $B$-index and EI-index highlight a tendency for external integration and information balance within the contralateral hemisphere (Fig.~\ref{fig4}), these metrics primarily capture the interactive dynamics between the hemispheres. A complementary analysis is therefore required to determine how the hemispheres differ across general, network-wide measures.
     
\begin{figure}  
     	\centering
     	\includegraphics[width=0.38\textwidth]{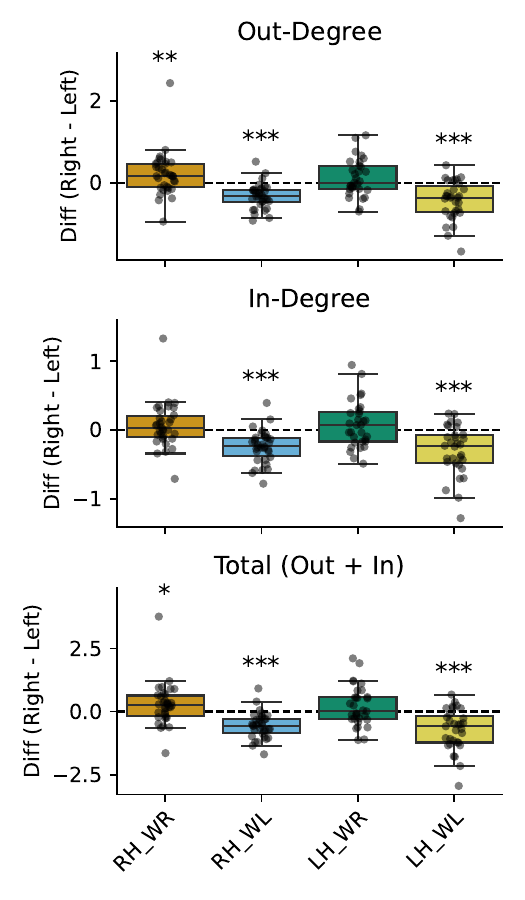}
     	\caption{\textbf{Subject-level distributions of hemispheric connectivity differences.} The plots display the individual hemispheric differences (Right Hemisphere minus Left Hemisphere) for out-degree, in-degree, and total degree across all group-task combinations. Statistical significance was assessed using pairwise Wilcoxon tests (*p < 0.05, **p < 0.01, ***p < 0.001). Boxplot bars represent 1.5 Interquartile Range (IQR).}
     	\label{fig5}
     \end{figure}

     Evaluating the overall difference between the right and left hemispheres revealed a clear ipsilateral dominance in tasks played by the left hand (favoring the hemisphere on the same side as the active hand) for out-degree, in-degree, and total degree. This significance was confirmed using a Wilcoxon paired test. With the single exception of left-handers writing with their right hand, all interhemispheric comparisons were significantly different and consistently characterized by this ipsilateral dominance (Fig.~\ref{fig5}).
     
	\section{Discussion and conclusion}

	Consistent with other measures of cortical behavior during handwriting tasks~\cite{RAMOS2025103425,RAMOS2025117412}, PTE highlights the left central, right parietal, and occipital regions, which correspond, respectively, to the motor control, impedance control, and visual feedback inherent to the task (see Figs.~\ref{Fig2} and~\ref{Fig3}). The results point to an alignment between right- and left-handers when using the same anatomical hand, both in hemispheric activity and interhemispheric interactions, rather than a mirrored response. Such mirroring may occur during motor imagery or simple tasks, where the task can still be controlled through the conventional motor system associated with the dominant hand. However, our findings show that during a fine motor task performed with the right hand, left-handers exhibit a inter-hemispheric pattern that is similar to right-handers.
	
	Our results offer a new perspective on brain lateralization during fine motor tasks. While previous research links motor task complexity to increased ipsilateral activation \cite{GAO20111280, WILKINS2020116344}, using a directional information approach we demonstrate that the direction of lateralization is highly scale-dependent for both left- and right-handers: ipsilateral dominance occurs primarily in general network-wide interactions (Fig.~\ref{fig5}), whereas inter-hemispheric interactions maintain a contralateral profile (Fig.~\ref{fig4}).
	
	Ultimately, this work demonstrates that the brains of left-handers are functionally much more similar to those of right-handers than previously assumed, displaying highly consistent anatomically driven responses even in tasks requiring a high level of motor control that subjects perform only with their dominant hand. This advances our understanding of human motor evolution by suggesting that left-handers do not represent an anomalous group atypical motor control. Instead, they possess an equivalent functional brain hemispherical information flow, with their distinct hand preference likely stemming from epigenetic factors~\cite{ocklenburg_genetics_2025,https://doi.org/10.1111/psyp.14676,https://doi.org/10.1002/dev.70143}. Conclusively, brain lateralization may be understood as an adaptive process optimizing single-handed tool manipulation. However, the human brain retains sufficient plasticity to specialize the opposite hand under certain conditions.
	
	\section{Methods}
	
	\subsection{Permutation Transfer Entropy (PTE)}
	
	Permutation transfer entropy (PTE)~\cite{PhysRevLett.100.158101} is an information-theoretic measure that quantifies directed statistical dependencies between time series based on their temporal ordering rather than their absolute amplitude. By encoding the signals as ordinal patterns, PTE is less sensitive to amplitude scaling, baseline differences, and recording-related variability, such as differences in electrode impedance or signal magnitude across subjects. This is particularly relevant for EEG, where amplitude variations do not necessarily reflect changes in the underlying temporal dynamics~\cite{e14081553,ZUNINO20173627,lehnertz_ordinal_2023}. Instead, the ordinal representation emphasizes the sequential structure of the signals, allowing PTE to characterize how the dynamics of one electrode contribute to the predictability of another~\cite{RAMOS2025117412}.
	
	In contrast to conventional synchronization measures, which primarily quantify the strength of coupling between signals, PTE also captures its directionality. Therefore, PTE provides information about asymmetric information transfer between electrodes while reducing the dependence of the measure on absolute signal amplitude.
	
	PTE bridges the concepts of standard Transfer Entropy (TE)~\cite{PhysRevLett.85.461}, which measures the predictive asymmetry between two signals, and Permutation Entropy (PE)~\cite{PhysRevLett.88.174102}, which maps continuous time series into discrete ordinal patterns. This combination yields an information-theoretic measure that is highly robust to dynamic noise and invariant to monotonic non-linear transformations.
	
	Let $X = \{x_t\}_{t=1}^N$ and $Y = \{y_t\}_{t=1}^N$ represent two interacting time series. We first reconstruct the phase space of both series using an embedding dimension $D$ (which dictates the length of the ordinal patterns).For each time step t, the D-dimensional vectors are defined as motifs~\cite{ROSARIO20157}, namely $\mathbf{x}_t^{D} = (x_t, x_{t+1}, \dots, x_{t+(D-1)})$ and similarly for $\mathbf{y}_t^{D}$, are mapped to their corresponding ordinal patterns (motifs), denoted as $\pi^X_t$ and $\pi^Y_t$. These symbols represent the relative sorting (ranking) of the values within the embedding window.
	
	D determines the number of possible detectable motifs. For the probability distribution of motifs to be well-defined, the motif sequence must contain a sufficient number of frames (T) for each embedding dimension, satisfying: D! $\times$ 5 < T~\cite{e21040385}.
	
	We used 60 seconds of data sampled at 1000 Hz and D=5. By considering non-overlapping motifs to avoid biasing the predictability of the subsequent state, we obtained 12,000 motifs. This provides a sequence length that is sufficiently large for the selected embedding dimensions and ensures the numerical convergence of the probability distribution.

	Transfer entropy from \(X\) to \(Y\) quantifies the reduction in uncertainty about the future ordinal pattern of \(Y\) obtained by additionally knowing the previous motif of \(X\), beyond the information contained in the previous motif of \(Y\). The permutation transfer entropy (PTE) from \(X\) to \(Y\) is defined as~\cite{PhysRevLett.100.158101}:
    \begin{align*}
PTE_{X \to Y}=\\
\sum_{\pi^Y_{t+1},,\pi^Y_t,,\pi^X_t}
p(\pi^Y_{t+1},\pi^Y_t,\pi^X_t)
\log\left(
\frac{
p(\pi^Y_{t+1}\mid\pi^Y_t,\pi^X_t)
}{
p(\pi^Y_{t+1}\mid\pi^Y_t)
}
\right),
   \end{align*}
   where \(p(\cdot\mid\cdot)\) denotes a conditional probability, and the probabilities are estimated from the relative frequencies of the corresponding motifs and their joint occurrences in the time series.

	A value of $PTE_{X \to Y} > 0$ indicates that the ordinal temporal structure of $X$ contains information that improves the prediction of the future state transitions of $Y$, beyond what is already predictable from $Y$'s own past. Therefore, it serves as a measure of predictive causality. In practice, evaluating both $PTE_{X \to Y}$ and $PTE_{Y \to X}$ allows us to determine the dominant direction of information flow and establish which variable is the driver and which is the responder in a complex dynamical system.
	
	By normalizing the PTE by the total information available in the target signal, we obtain the fraction of the target information that is influenced by the source:$$norm(PTE_{X \to Y}) = \frac{PTE_{X \to Y}}{H(Y)}$$where $H(Y)$ is the PE of the target electrode:$$H(Y) = - \sum_{\pi^Y} p(\pi^Y) \log p(\pi^Y)$$and $p(\pi^Y)$ denotes the marginal probability of encountering the motif $\pi^Y$ across the entire time series.

	\subsection{EEG preprocessing}

	EEG (10–20 cap system) recordings were acquired under controlled experimental conditions, with electrode impedances maintained below $50k\Omega$ throughout the recordings. 
	To specifically focus on hemispheric lateralization, the 64 channel EEG montage was restricted to electrodes with anatomically corresponding positions in the left and right hemispheres. The reference channel (Cz), Midline electrodes (e.g. Fz, and Pz) and non-cortical electrodes (e.g., M1, M2, FT11, F11, F12 and FT12) were excluded. This procedure resulted in a set of 48 electrodes, comprising 24 left-hemisphere electrodes and their 24 corresponding right-hemisphere electrodes ($\mathcal{E}_L \cup \mathcal{E}_R$), where $\mathcal{E}_
	L$ and $\mathcal{E}_R$ denote the sets of left- and right-hemisphere electrodes, respectively.
	
	The selected electrode pairs ($\mathcal{E}_L / \mathcal{E}_R$) were: (FP1/FP2), (AF3/AF4), (F7/F8), (F5/F6), (F3/F4), (FC5/FC6), (FC3/FC4), (FC1/FC2), (F1/F2), (C5/C6), (C3/C4), (C1/C2), (TP7/TP8), (T7/T8), (CP5/CP6), (CP3/CP4), (CP1/CP2), (P7/P8), (P5/P6), (P3/P4), (P1/P2), (PO3/PO4), (O1/O2), and (CB1/CB2).
		
	No band-pass filtering was applied, as filtering can modify the temporal structure of the signal~\cite{10.3389/fninf.2026.1791461} and introduce edge/transient effects that may affect the motif representation used for PTE.
	 
 	To remove muscle artifacts and potential electrode conduction noise, the signals were segmented into 1s epochs. Epochs in which any valid electrode has the absolute amplitude exceeded 300 $\mu$V was discarded, and the retained epochs were then concatenated in order. This 1s window size was selected to minimize temporal discontinuities, given that PTE is a metric highly dependent on sequential data structure. With a sampling rate of 1000 Hz and an embedding dimension of $D=5$, each 1s epoch contains 200 non-overlapping motifs. In the worst-case scenario, the concatenation process introduces 59 epoch boundaries where the temporal sequence is compromised. This disruption affects less than 0.5$\%$ of the total motifs; therefore, because PTE is a cumulative measure, we consider the impact of these discontinuities to be negligible.

	 \subsection{Pairwise matrix similarity and correlation metrics}
	 
     To quantify the intra-subject differences in neural organization between tasks performed with the right and left hands, we applied three metrics to the individual PTE adjacency matrices and their corresponding hierarchical structures:  
	 
	 \textbf{Mean Absolute Error (MAE)}
	We used the MAE to evaluate the global absolute magnitude of difference between two matrices. For an individual's right-hand connectivity matrix $X$ and left-hand connectivity matrix $Y$, each of size $n \times n$, the MAE computes the average cell-by-cell absolute difference:

	 $$MAE = \frac{1}{N} \sum_{i=1}^{n} \sum_{j=1}^{n} \vert{}X_{ij} - Y_{ij}\vert{}$$

	 where $N$ is the total number of electrodes.
	 
	 \textbf{Mantel Correlation}
	To assess the structural association between the functional networks, we applied the Mantel test. This method is specifically designed to evaluate the linear relationship between the corresponding elements of two distance or similarity matrices, determining if the overall network architecture remains conserved across hands. The Mantel correlation coefficient $r$ is defined as:

	 $$r = \frac{1}{N-1} \sum_{i \neq j} \left( \frac{X_{ij} - \bar{X}}{s_X} \right) \left( \frac{Y_{ij} - \bar{Y}}{s_Y} \right)$$

	where $\bar{X}$ and $\bar{Y}$ are the mean values of the off-diagonal elements for matrices $X$ and $Y$, respectively, and $s_X$ and $s_Y$ are their standard deviations. Statistical significance was assessed using matrix permutations.
	 
	 \textbf{Cophenetic Correlation Coefficient}
	To evaluate how consistently the macroscopic topology of information flow is preserved, we quantified the similarity between the hierarchical clustering patterns (dendrograms) generated for each hand using the cophenetic correlation coefficient. This index computes the linear correlation between the original pairwise distance values and the cophenetic distances (the tree height at which two specific electrodes are first joined in the dendrogram). It is calculated as:

	 $$c = \frac{\sum_{i<j} (x_{ij} - \bar{x})(c_{ij} - \bar{c})}{\sqrt{\sum_{i<j} (x_{ij} - \bar{x})^2 \sum_{i<j} (c_{ij} - \bar{c})^2}}$$

	where $x_{ij}$ is the distance between electrodes $i$ and $j$ in the original distance matrix, $c_{ij}$ is their cophenetic distance in the dendrogram model, and $\bar{x}$ and $\bar{c}$ are the respective averages of these distances.

	\subsection{Inter-hemispheric connectivity metrics}
	For each subject and task, brain PTE was represented by a directed adjacency matrix (PTE), where $PTE_{ij}$ denotes a information originating from channel (i) and directed toward channel (j). For each node $\mathcal{E}$, the in-degree ($k_{\mathcal{E}}^{\mathrm{in}}$), out-degree ($k_{\mathcal{E}}^{\mathrm{out}}$), and total degree ($k_{\mathcal{E}}$)  were calculated as:
	
	\begin{equation*}
		k_{\mathcal{E}}^{\mathrm{out}} = \sum_j PTE_{\mathcal{E},j},
		\qquad
		k_{\mathcal{E}}^{\mathrm{in}} = \sum_i PTE_{i,\mathcal{E}}
	\end{equation*}
	
	\begin{equation*}
		k_{\mathcal{E}} = k_{\mathcal{E}}^{\mathrm{in}} + k_{\mathcal{E}}^{\mathrm{out}}.
	\end{equation*}
	
	To characterize the directionality of connections between the two brain hemispheres, the external-internal (EI) index was calculated separately for each source hemisphere (H) as~\cite{toutain_brain_2022}:
	
	\begin{equation*}
		EI_H = \frac{E_H-I_H}{E_H+I_H},
	\end{equation*}
	
	where (E) represents the total connectivity originating from the source hemisphere and directed toward the opposite hemisphere, whereas (I) represents the total connectivity originating from the source hemisphere and remaining within the same hemisphere. Thus, (EI>0) indicates a predominance of interhemispheric connections, whereas (EI<0) indicates predominantly intrahemispheric connectivity.
	
	The directional balance of connectivity was additionally quantified using the flow balance index by hemisphere:

	\begin{equation*}
		Out_H = \sum_{e\in \mathcal{E}_H} k_e^{\mathrm{out}},
		\qquad
		In_H = \sum_{e\in \mathcal{E}_H} k_e^{\mathrm{in}}
	\end{equation*}
	
	\begin{equation*}
		B_H = \frac{Out_H-In_H}{Out_H+In_H},
	\end{equation*}
	
	where (Out) and (In) represent the total outgoing and incoming transfer entropy of the considered hemisphere, respectively. Positive values indicate a predominance of outgoing flow, whereas negative values indicate a predominance of incoming flow.

	\section*{Ethics approval}
	The experimental protocol was approved by the ethical committee of the Farmacy Faculty of Federal University of Bahia (Certificate of Presentation for Ethical Appreciation: 68289021.5.0000.5531) in accordance with the Declaration of Helsinki.
	
	\section*{Acknowledgments}
	We thank the Brazilian Federal Agency for Support and Evaluation of Graduate Education (CAPES) for supporting Yago Emanoel Ramos through grant no. 88887.203818/2025-00. We thank Henrique Ferraz de Arruda from University of Zaragoza for valuable suggestions and methodological discussions.
	
	\bibliography{ref}

@article{FERNANDES2027116449,
title = {Manual asymmetries and cortical modulation: relationships between motor control and interhemispheric connectivity},
journal = {Behavioural Brain Research},
volume = {516},
pages = {116449},
year = {2027},
issn = {0166-4328},
doi = {https://doi.org/10.1016/j.bbr.2026.116449},
url = {https://www.sciencedirect.com/science/article/pii/S0166432826004250},
author = {Lidiane Aparecida Fernandes and Tércio Apolinário-Souza and Lucas Eduardo Antunes Bicalho and Marco Antonio Cavalcanti Garcia and Guilherme Menezes Lage}
}

@article{feng_multimodal_2026,
	title = {A {Multimodal} {fNIRS}–{EEG} {Dataset} for {Unilateral} {Limb} {Motor} {Imagery}},
	volume = {13},
	issn = {2052-4463},
	url = {https://doi.org/10.1038/s41597-026-07807-x},
	doi = {10.1038/s41597-026-07807-x},
	number = {1},
	journal = {Scientific Data},
	author = {Feng, Lufeng and Xu, Baomin and Zhang, Haoran and Lin, Bihai and Deng, Zuxuan and Tao, Sidi and Liu, Chenyu and Jia, Shifan and Duan, Li and Jia, Ziyu},
	month = aug,
	year = {2026},
	pages = {1159},
}

@article{Wiper02112017,
	author = {Mallory L. Wiper},
	title = {Evolutionary and mechanistic drivers of laterality: A review and new synthesis},
	journal = {Laterality},
	volume = {22},
	number = {6},
	pages = {740--770},
	year = {2017},
	publisher = {Routledge},
	doi = {10.1080/1357650X.2017.1291658},
	
	note ={PMID: 28276877},
	
	
	URL = { 
	
	https://doi.org/10.1080/1357650X.2017.1291658
	
	
	
	},
	eprint = { 
	
	https://doi.org/10.1080/1357650X.2017.1291658
	
	
	
	}
	
}

@article{blum_animal_2018,
	title = {Animal left–right asymmetry},
	volume = {28},
	issn = {0960-9822},
	url = {https://doi.org/10.1016/j.cub.2018.02.073},
	doi = {10.1016/j.cub.2018.02.073},
	number = {7},
	urldate = {2026-08-15},
	journal = {Current Biology},
	author = {Blum, Martin and Ott, Tim},
	month = apr,
	year = {2018},
	note = {Publisher: Elsevier},
	pages = {R301--R304},
}

@Inbook{Verendeev2016,
	author="Verendeev, Andrey
	and Sherwood, Chet C.
	and Hopkins, William D.",
	editor="Kivell, Tracy L.
	and Lemelin, Pierre
	and Richmond, Brian G.
	and Schmitt, Daniel",
	title="Organization and Evolution of the Neural Control of the Hand in Primates: Motor Systems, Sensory Feedback, and Laterality",
	bookTitle="The Evolution of the Primate Hand: Anatomical, Developmental, Functional, and Paleontological Evidence",
	year="2016",
	publisher="Springer New York",
	address="New York, NY",
	pages="131--153",
	isbn="978-1-4939-3646-5",
	doi="10.1007/978-1-4939-3646-5_6",
	url="https://doi.org/10.1007/978-1-4939-3646-5_6"
}

@article {10.7554/eLife.77875,
	article_type = {journal},
	title = {The evolution and biological correlates of hand preferences in anthropoid primates},
	author = {Caspar, Kai R and Pallasdies, Fabian and Mader, Larissa and Sartorelli, Heitor and Begall, Sabine},
	editor = {Tung, Jenny and Przeworski, Molly and DeCasien, Alex and Meguerditchian, Adrien},
	volume = 11,
	year = 2022,
	month = {dec},
	pub_date = {2022-12-01},
	pages = {e77875},
	citation = {eLife 2022;11:e77875},
	doi = {10.7554/eLife.77875},
	url = {https://doi.org/10.7554/eLife.77875},
	journal = {eLife},
	issn = {2050-084X},
	publisher = {eLife Sciences Publications, Ltd},
}

@article{
	Ocklenburg2022,
	author = {Sebastian Ocklenburg  and Annakarina Mundorf },
	title = {Symmetry and asymmetry in biological structures},
	journal = {Proceedings of the National Academy of Sciences},
	volume = {119},
	number = {28},
	pages = {e2204881119},
	year = {2022},
	doi = {10.1073/pnas.2204881119},
	URL = {https://www.pnas.org/doi/abs/10.1073/pnas.2204881119},
	eprint = {https://www.pnas.org/doi/pdf/10.1073/pnas.2204881119}}

@article{renfrew_neural_2008,
	title = {Neural correlates of {Early} {Stone} {Age} toolmaking: technology, language and cognition in human evolution},
	volume = {363},
	issn = {0962-8436},
	url = {https://doi.org/10.1098/rstb.2008.0001},
	doi = {10.1098/rstb.2008.0001},
	number = {1499},
	journal = {Philosophical Transactions of the Royal Society B: Biological Sciences},
	author = {Renfrew, Colin and Frith, Chris and Malafouris, Lambros and Stout, Dietrich and Toth, Nicholas and Schick, Kathy and Chaminade, Thierry},
	month = feb,
	year = {2008},
	pages = {1939--1949},
}

@article{https://doi.org/10.1111/eth.12827,
	author = {Prieur, Jacques and Lemasson, Alban and Barbu, Stéphanie and Blois-Heulin, Catherine},
	title = {History, development and current advances concerning the evolutionary roots of human right-handedness and language: Brain lateralisation and manual laterality in non-human primates},
	journal = {Ethology},
	volume = {125},
	number = {1},
	pages = {1-28},
	doi = {https://doi.org/10.1111/eth.12827},
	url = {https://onlinelibrary.wiley.com/doi/abs/10.1111/eth.12827},
	eprint = {https://onlinelibrary.wiley.com/doi/pdf/10.1111/eth.12827},
	year = {2019}
}

@article{sainburg_evidence_2002,
	title = {Evidence for a dynamic-dominance hypothesis of handedness},
	volume = {142},
	issn = {1432-1106},
	url = {https://doi.org/10.1007/s00221-001-0913-8},
	doi = {10.1007/s00221-001-0913-8},
	number = {2},
	journal = {Experimental Brain Research},
	author = {Sainburg, Robert L.},
	month = jan,
	year = {2002},
	pages = {241--258},
}

@ARTICLE{10.3389/fpsyg.2014.01092,
	
	AUTHOR={Sainburg, Robert L. },
	
	TITLE={Convergent models of handedness and brain lateralization},
	
	JOURNAL={Frontiers in Psychology},
	
	VOLUME={Volume 5 - 2014},
	
	YEAR={2014},
	
	URL={https://www.frontiersin.org/journals/psychology/articles/10.3389/fpsyg.2014.01092},
	
	DOI={10.3389/fpsyg.2014.01092},
	
	ISSN={1664-1078}}

@article{YADAV2014385,
	title = {Handedness can be explained by a serial hybrid control scheme},
	journal = {Neuroscience},
	volume = {278},
	pages = {385-396},
	year = {2014},
	issn = {0306-4522},
	doi = {https://doi.org/10.1016/j.neuroscience.2014.08.026},
	url = {https://www.sciencedirect.com/science/article/pii/S0306452214007039},
	author = {V. Yadav and R.L. Sainburg}
}

@ARTICLE{10.3389/fpsyg.2017.01021,
	
	AUTHOR={Corballis, Michael C. },
	
	TITLE={The Evolution of Lateralized Brain Circuits},
	
	JOURNAL={Frontiers in Psychology},
	
	VOLUME={Volume 8 - 2017},
	
	YEAR={2017},
	
	URL={https://www.frontiersin.org/journals/psychology/articles/10.3389/fpsyg.2017.01021},
	
	DOI={10.3389/fpsyg.2017.01021},
	
	ISSN={1664-1078}}

@article{corballis_evolution_2008,
	title = {The evolution and genetics of cerebral asymmetry},
	volume = {364},
	issn = {0962-8436},
	url = {https://doi.org/10.1098/rstb.2008.0232},
	doi = {10.1098/rstb.2008.0232},
	number = {1519},
	journal = {Philosophical Transactions of the Royal Society B: Biological Sciences},
	author = {Corballis, Michael C},
	month = dec,
	year = {2008},
	pages = {867--879},
}

@article{
	Sha2021,
	author = {Zhiqiang Sha  and Antonietta Pepe  and Dick Schijven  and Amaia Carrión-Castillo  and James M. Roe  and René Westerhausen  and Marc Joliot  and Simon E. Fisher  and Fabrice Crivello  and Clyde Francks },
	title = {Handedness and its genetic influences are associated with structural asymmetries of the cerebral cortex in 31,864 individuals},
	journal = {Proceedings of the National Academy of Sciences},
	volume = {118},
	number = {47},
	pages = {e2113095118},
	year = {2021},
	doi = {10.1073/pnas.2113095118},
	URL = {https://www.pnas.org/doi/abs/10.1073/pnas.2113095118},
	eprint = {https://www.pnas.org/doi/pdf/10.1073/pnas.2113095118}}

@article{hagemann_advantage_2009,
	title = {The advantage of being left-handed in interactive sports},
	volume = {71},
	issn = {1943-393X},
	url = {https://doi.org/10.3758/APP.71.7.1641},
	doi = {10.3758/APP.71.7.1641},
	number = {7},
	journal = {Attention, Perception, \& Psychophysics},
	author = {Hagemann, Norbert},
	month = oct,
	year = {2009},
	pages = {1641--1648},
}

@article{loffing_left-handedness_2017,
	title = {Left-handedness and time pressure in elite interactive ball games},
	volume = {13},
	issn = {1744-9561},
	url = {https://doi.org/10.1098/rsbl.2017.0446},
	doi = {10.1098/rsbl.2017.0446},
	number = {11},
	journal = {Biology Letters},
	author = {Loffing, Florian},
	month = nov,
	year = {2017},
	pages = {20170446},
}

@article{simon_prevalence_2025,
	title = {Prevalence of left-handers and their role in antagonistic sports: beyond mere counts towards a more in-depth distributional analysis of ranking data},
	volume = {12},
	issn = {2054-5703},
	url = {https://doi.org/10.1098/rsos.250303},
	doi = {10.1098/rsos.250303},
	number = {9},
	journal = {Royal Society Open Science},
	author = {Simon, Tim and Loffing, Florian and Frasnelli, Elisa},
	month = sep,
	year = {2025},
	pages = {250303},
}

@article{BADZAKOVATRAJKOV20103086,
	title = {Cerebral asymmetries in monozygotic twins: An fMRI study},
	journal = {Neuropsychologia},
	volume = {48},
	number = {10},
	pages = {3086-3093},
	year = {2010},
	issn = {0028-3932},
	doi = {https://doi.org/10.1016/j.neuropsychologia.2010.06.020},
	url = {https://www.sciencedirect.com/science/article/pii/S0028393210002599},
	author = {Gjurgjica Badzakova-Trajkov and Isabelle S. Häberling and Michael C. Corballis}
}

@article{steinmetz_brain_1995,
	title = {Brain ({A}){Symmetry} in {Monozygotic} {Twins}},
	volume = {5},
	issn = {1047-3211},
	url = {https://doi.org/10.1093/cercor/5.4.296},
	doi = {10.1093/cercor/5.4.296},
	number = {4},
	journal = {Cerebral Cortex},
	author = {Steinmetz, Helmuth and Herzog, Axel and Schlaug, Gottfried and Huang, Yanxiong and Jäncke, Lutz},
	month = jul,
	year = {1995},
	pages = {296--300},
}

@article{kloppel2007can,
	title={Can left-handedness be switched? Insights from an early switch of handwriting},
	author={Kl{\"o}ppel, Stefan and Vongerichten, Anna and Eimeren, Thilo van and Frackowiak, Richard SJ and Siebner, Hartwig R},
	doi={https://doi.org/10.1523/JNEUROSCI.1299-07.2007},
	journal={The Journal of Neuroscience},
	volume={27},
	number={29},
	pages={7847--7853},
	year={2007},
	publisher={Society for Neuroscience}
}

@article{KARLSSON2024108837,
	title = {Hemispheric asymmetry of hand and tool perception in left- and right-handers with known language dominance},
	journal = {Neuropsychologia},
	volume = {196},
	pages = {108837},
	year = {2024},
	issn = {0028-3932},
	doi = {https://doi.org/10.1016/j.neuropsychologia.2024.108837},
	url = {https://www.sciencedirect.com/science/article/pii/S0028393224000526},
	author = {Emma M. Karlsson and David P. Carey}
}

@article{TZOURIOMAZOYER2016319,
	title = {The neural bases of hemispheric specialization},
	journal = {Neuropsychologia},
	volume = {93},
	pages = {319-324},
	year = {2016},
	note = {The Neural Bases of Hemispheric Specialisation},
	issn = {0028-3932},
	doi = {https://doi.org/10.1016/j.neuropsychologia.2016.10.010},
	url = {https://www.sciencedirect.com/science/article/pii/S0028393216303840},
	author = {Nathalie Tzourio-Mazoyer and Mohamed L. Seghier}
}

@ARTICLE{10.3389/neuro.09.039.2009,
	
	AUTHOR={Willems, Roel M. and Toni, Ivan  and Hagoort, Peter  and Casasanto, Daniel },
	
	TITLE={Body-specific motor imagery of hand actions: neural evidence from right- and left-handers},
	
	JOURNAL={Frontiers in Human Neuroscience},
	
	VOLUME={Volume 3 - 2009},
	
	YEAR={2009},
	
	URL={https://www.frontiersin.org/journals/human-neuroscience/articles/10.3389/neuro.09.039.2009},
	
	DOI={10.3389/neuro.09.039.2009},
	
	ISSN={1662-5161}}

@article{10.1371/journal.pone.0036036,
		doi = {10.1371/journal.pone.0036036},
		author = {Brookshire, Geoffrey AND Casasanto, Daniel},
		journal = {PLOS ONE},
		publisher = {Public Library of Science},
		title = {Motivation and Motor Control: Hemispheric Specialization for Approach Motivation Reverses with Handedness},
		year = {2012},
		month = {04},
		volume = {7},
		url = {https://doi.org/10.1371/journal.pone.0036036},
		pages = {1-5},
		number = {4},
		
	}

@article{RAMOS2025103425,
	title = {Handedness and brain lateralization: A nonlinear motor approach combined with EEG},
	journal = {Human Movement Science},
	volume = {104},
	pages = {103425},
	year = {2025},
	issn = {0167-9457},
	doi = {https://doi.org/10.1016/j.humov.2025.103425},
	url = {https://www.sciencedirect.com/science/article/pii/S0167945725001071},
	author = {Yago Emanoel Ramos and Mariana Teixeira Santos and Iago Nudelman Reis Yamamoto and Cecília Bastos da Costa Accioly and Jean-François Daneault and Daniel Gomes {de Almeida Filho} and José Garcia Vivas Miranda}
}

@article{RAMOS2025117412,
	title = {Linking biomechanical model dynamics and neural complexity: Permutation entropy approaches to motor control},
	journal = {Chaos, Solitons $\&$ Fractals},
	volume = {201},
	pages = {117412},
	year = {2025},
	issn = {0960-0779},
	doi = {https://doi.org/10.1016/j.chaos.2025.117412},
	url = {https://www.sciencedirect.com/science/article/pii/S0960077925014250},
	author = {Yago Emanoel Ramos and Ângelo Frederico Torres and Cecília Bastos {da Costa Accioly} and Fernanda Selingardi Matias and José Garcia Vivas Miranda},
	
}

@article{RAMOS2026118081,
	title = {Emergent togetherness through multilayer and high-order synchronization in generative dance neuro-motor systems},
	journal = {Chaos, Solitons $\&$ Fractals},
	volume = {208},
	pages = {118081},
	year = {2026},
	issn = {0960-0779},
	doi = {https://doi.org/10.1016/j.chaos.2026.118081},
	url = {https://www.sciencedirect.com/science/article/pii/S0960077926002225},
	author = {Yago Emanoel Ramos and Raphael Silva {do Rosário} and Adriana {de Faria Gehres} and Maria João Alves and Ana Maria Leitão and Cecília Bastos {da Costa Accioly} and Fatima Wachowicz and Ivani Lúcia Oliveira {de Santana} and José Garcia Vivas Miranda}
}

@article{doi:10.1177/15500594251394773,
	author = {Kleber Lopes Lima Fialho and José Garcia Vivas Miranda and Yago Emanoel Ramos and Rita de Cássia Saldanha de Lucena},
	title ={Characterization of Neurophysiological, Motor, and Emotional Biomarkers in Adolescents with ASD: An Integrated Analysis with qEEG, Facial Expression, and Biomechanics Analysis},
	
	journal = {Clinical EEG and Neuroscience},
	volume = {57},
	number = {2},
	pages = {141-151},
	year = {2026},
	doi = {10.1177/15500594251394773},
	note ={PMID: 41264522},
	
	URL = { 
	
	https://doi.org/10.1177/15500594251394773
	
	
	
	},
	eprint = { 
	
	https://doi.org/10.1177/15500594251394773
	
	
	
	}
}

@article{ocklenburg_genetics_2025,
	title = {Genetics of human handedness: microtubules and beyond},
	volume = {41},
	issn = {0168-9525},
	url = {https://doi.org/10.1016/j.tig.2025.01.006},
	doi = {10.1016/j.tig.2025.01.006},
	number = {6},
	urldate = {2026-08-19},
	journal = {Trends in Genetics},
	author = {Ocklenburg, Sebastian and Mundorf, Annakarina and Peterburs, Jutta and Paracchini, Silvia},
	month = jun,
	year = {2025},
	note = {Publisher: Elsevier},
	pages = {497--505},
}

@article{williams_structural_2023,
	title = {Structural and functional asymmetry of the neonatal cerebral cortex},
	volume = {7},
	issn = {2397-3374},
	url = {https://doi.org/10.1038/s41562-023-01542-8},
	doi = {10.1038/s41562-023-01542-8},
	number = {6},
	journal = {Nature Human Behaviour},
	author = {Williams, Logan Z. J. and Fitzgibbon, Sean P. and Bozek, Jelena and Winkler, Anderson M. and Dimitrova, Ralica and Poppe, Tanya and Schuh, Andreas and Makropoulos, Antonios and Cupitt, John and O’Muircheartaigh, Jonathan and Duff, Eugene P. and Cordero-Grande, Lucilio and Price, Anthony N. and Hajnal, Joseph V. and Rueckert, Daniel and Smith, Stephen M. and Edwards, A. David and Robinson, Emma C.},
	month = jun,
	year = {2023},
	pages = {942--955},
}

@article{PhysRevLett.85.461,
	title = {Measuring Information Transfer},
	author = {Schreiber, Thomas},
	journal = {Phys. Rev. Lett.},
	volume = {85},
	issue = {2},
	pages = {461--464},
	numpages = {0},
	year = {2000},
	month = {Jul},
	publisher = {American Physical Society},
	doi = {10.1103/PhysRevLett.85.461},
	url = {https://link.aps.org/doi/10.1103/PhysRevLett.85.461}
}

@article{PhysRevLett.88.174102,
	title = {Permutation Entropy: A Natural Complexity Measure for Time Series},
	author = {Bandt, Christoph and Pompe, Bernd},
	journal = {Phys. Rev. Lett.},
	volume = {88},
	issue = {17},
	pages = {174102},
	numpages = {4},
	year = {2002},
	month = {Apr},
	publisher = {American Physical Society},
	doi = {10.1103/PhysRevLett.88.174102},
	url = {https://link.aps.org/doi/10.1103/PhysRevLett.88.174102}
}

@article{PhysRevLett.100.158101,
	title = {Symbolic Transfer Entropy},
	author = {Staniek, Matth\"aus and Lehnertz, Klaus},
	journal = {Phys. Rev. Lett.},
	volume = {100},
	issue = {15},
	pages = {158101},
	numpages = {4},
	year = {2008},
	month = {Apr},
	publisher = {American Physical Society},
	doi = {10.1103/PhysRevLett.100.158101},
	url = {https://link.aps.org/doi/10.1103/PhysRevLett.100.158101}
}

@Article{e14081553,
	AUTHOR = {Zanin, Massimiliano and Zunino, Luciano and Rosso, Osvaldo A. and Papo, David},
	TITLE = {Permutation Entropy and Its Main Biomedical and Econophysics Applications: A Review},
	JOURNAL = {Entropy},
	VOLUME = {14},
	YEAR = {2012},
	NUMBER = {8},
	PAGES = {1553--1577},
	URL = {https://www.mdpi.com/1099-4300/14/8/1553},
	ISSN = {1099-4300},
	DOI = {10.3390/e14081553}
}

@article{ZUNINO20173627,
	title = {Detecting nonlinearity in short and noisy time series using the permutation entropy},
	journal = {Physics Letters A},
	volume = {381},
	number = {42},
	pages = {3627-3635},
	year = {2017},
	issn = {0375-9601},
	doi = {https://doi.org/10.1016/j.physleta.2017.09.032},
	url = {https://www.sciencedirect.com/science/article/pii/S0375960117308976},
	author = {Luciano Zunino and Christopher W. Kulp}
}

@article{lehnertz_ordinal_2023,
	title = {Ordinal methods for a characterization of evolving functional brain networks},
	volume = {33},
	issn = {1054-1500},
	url = {https://doi.org/10.1063/5.0136181},
	doi = {10.1063/5.0136181},
	number = {2},
	journal = {Chaos: An Interdisciplinary Journal of Nonlinear Science},
	author = {Lehnertz, Klaus},
	month = feb,
	year = {2023},
	pages = {022101},
}

@article{toutain_brain_2022,
	title = {Brain {Asymmetry} in {Pain} {Affective} {Modulation}},
	volume = {23},
	issn = {1526-4637},
	url = {https://doi.org/10.1093/pm/pnab232},
	doi = {10.1093/pm/pnab232},
	number = {4},
	journal = {Pain Medicine},
	author = {Toutain, Thaise Graziele L de O and Alba, Guzmán and Miranda, José Garcia Vivas and Silva do Rosário, Raphael and Muñoz, Miguel and de Sena, Eduardo Pondé},
	month = apr,
	year = {2022},
	pages = {686--696},
}

@article{https://doi.org/10.1002/dev.70143,
	author = {Hamaoui, Jad and Zhang, Hao and King, Suzanne and Castellanos-Ryan, Natalie},
	title = {Prenatal Maternal Stress and Weak Handedness in Early Childhood: The Iowa Flood Study},
	journal = {Developmental Psychobiology},
	volume = {68},
	number = {2},
	pages = {e70143},
	doi = {https://doi.org/10.1002/dev.70143},
	url = {https://onlinelibrary.wiley.com/doi/abs/10.1002/dev.70143},
	eprint = {https://onlinelibrary.wiley.com/doi/pdf/10.1002/dev.70143},
	year = {2026}
}

@article{https://doi.org/10.1111/psyp.14676,
	author = {Hamaoui, Jad and Ocklenburg, Sebastian and Segond, Hervé},
	title = {Perinatal adversities as a common factor underlying the association between atypical laterality and neurodevelopmental disorders: A developmental perspective},
	journal = {Psychophysiology},
	volume = {61},
	number = {12},
	pages = {e14676},
	doi = {https://doi.org/10.1111/psyp.14676},
	url = {https://onlinelibrary.wiley.com/doi/abs/10.1111/psyp.14676},
	eprint = {https://onlinelibrary.wiley.com/doi/pdf/10.1111/psyp.14676},
	note = {e14676 PsyP-2024-0045.R2},
	year = {2024}
}

@article{GAO20111280,
	title = {Evaluation of effective connectivity of motor areas during motor imagery and execution using conditional Granger causality},
	journal = {NeuroImage},
	volume = {54},
	number = {2},
	pages = {1280-1288},
	year = {2011},
	issn = {1053-8119},
	doi = {https://doi.org/10.1016/j.neuroimage.2010.08.071},
	url = {https://www.sciencedirect.com/science/article/pii/S1053811910011687},
	author = {Qing Gao and Xujun Duan and Huafu Chen}
}

@article{WILKINS2020116344,
	title = {Coordination of multiple joints increases bilateral connectivity with ipsilateral sensorimotor cortices},
	journal = {NeuroImage},
	volume = {207},
	pages = {116344},
	year = {2020},
	issn = {1053-8119},
	doi = {https://doi.org/10.1016/j.neuroimage.2019.116344},
	url = {https://www.sciencedirect.com/science/article/pii/S1053811919309358},
	author = {Kevin B. Wilkins and Jun Yao}
}

@article{ROSARIO20157,
	title = {Motif-Synchronization: A new method for analysis of dynamic brain networks with EEG},
	journal = {Physica A: Statistical Mechanics and its Applications},
	volume = {439},
	pages = {7-19},
	year = {2015},
	issn = {0378-4371},
	doi = {https://doi.org/10.1016/j.physa.2015.07.018},
	url = {https://www.sciencedirect.com/science/article/pii/S0378437115006275},
	author = {R.S. Rosário and P.T. Cardoso and M.A. Muñoz and P. Montoya and J.G.V. Miranda}
}

@Article{e21040385,
	AUTHOR = {Cuesta-Frau, David and Murillo-Escobar, Juan Pablo and Orrego, Diana Alexandra and Delgado-Trejos, Edilson},
	TITLE = {Embedded Dimension and Time Series Length. Practical Influence on Permutation Entropy and Its Applications},
	JOURNAL = {Entropy},
	VOLUME = {21},
	YEAR = {2019},
	NUMBER = {4},
	ARTICLE-NUMBER = {385},
	URL = {https://www.mdpi.com/1099-4300/21/4/385},
	PubMedID = {33267099},
	ISSN = {1099-4300},
	DOI = {10.3390/e21040385}
}

@ARTICLE{10.3389/fninf.2026.1791461,
	
	AUTHOR={Wyczesany, Miroslaw  and Spadone, Sara  and Górski, Tomasz  and Łabaza, Adam  and Domagala, Michal  and Kaminski, Maciej  and Ligeza, Tomasz S.  and Capotosto, Paolo  and Kroker, Thomas  and Junghöfer, Markus  and Della Penna, Stefania },
	
	TITLE={ASCT: a pipeline for standardized analysis of MEG/EEG directional connectivity. Practical guidelines for applications of source-based Directed Transfer Function},
	
	JOURNAL={Frontiers in Neuroinformatics},
	
	VOLUME={Volume 20 - 2026},
	
	YEAR={2026},
	
	URL={https://www.frontiersin.org/journals/neuroinformatics/articles/10.3389/fninf.2026.1791461},
	
	DOI={10.3389/fninf.2026.1791461},
	
	ISSN={1662-5196}}
	
\end{document}